\documentclass[aps,prl,showpacs,onecolumn]{revtex4}
\usepackage{graphicx}
\usepackage[usenames]{color}
\pdfoutput=1
\def\bea{\begin{eqnarray}}
\def\eea{\end{eqnarray}}
\def\be{\begin{equation}}
\def\ee{\end{equation}}

\def\beq{\begin{equation}}
\def\eeq{\end{equation}}

\begin{document}
%\begin{center}
\title{\bf Generation of Higher-Order Harmonics By Addition of a High Frequency XUV Radiation to the IR One}

%April 22
%April 23

\author{Avner Fleischer and Nimrod Moiseyev}

\affiliation{Schulich Faculty of Chemistry and Minerva Center for
Nonlinear Physics of Complex Systems, Technion -- Israel Institute
of Technology, Haifa 32000,
Israel.}\email{avnerf@tx.technion.ac.il ,
nimrod@tx.technion.ac.il}

\date{\today}
\begin{abstract}
The irradiation of atoms by a strong IR laser field of frequency
$\omega$ results in the emission of odd-harmonics of $\omega$ ("IR
harmonics") up to some maximal cut-off frequency. The addition of an
XUV field of frequency $\tilde{q}\omega$ larger than the IR cut-off
frequency to the IR driver field leads to the appearance of new
higher-order harmonics ("XUV harmonics") $\tilde{q} \pm 2K,
2\tilde{q} \pm (2K-1), 3\tilde{q} \pm 2K,...$ ($K$ integer) which
were absent in the spectra in the presence of the IR field alone.
The mechanism responsible for the appearance of the XUV harmonics is
analyzed analytically using a generalization of the semiclassical
re-collision (three-step) model of high harmonic generation. It is
shown that the emitted HHG radiation field can be written as a serie
of terms, with the HHG field obtained from the three-step model in
its most familiar context [P. B. Corkum, \textit{Phys. Rev. Lett.}
{\bf 71}, 1994 (1993)] resulting from the zeroth-order term. The
origin of the higher-order terms is shown to be the ac-Stark
oscillations of the remaining ground electronic state which are
induced by the XUV field. These terms are responsible for the
appearance of the new XUV harmonics in the HGS. The XUV harmonics
are formed by the same electron trajectories which form the IR
harmonics and have the same emission times, but a much lower
intensity than the IR harmonics, due to the small quiver amplitude
of the ac-Stark oscillation. Nevertheless, this mechanism allows the
extension of the cut-off in the HGS without the necessity of
increasing the IR field intensity, as is verified by numerical
time-dependent Schr\"{o}dinger equation simulation of a Xe atom
shined by a combination of IR and XUV field.

\end{abstract}
\maketitle

%\pacs
{03.65.-w, 42.50.Hz, 42.65.-Ky, 32.80.Rm}

Focusing intense linearly-polarized monochromatic infra-red (IR)
laser pulses into gas of atoms can lead to the emission of
high-energy photons with frequencies extending into the extreme
ultraviolet (XUV) and X-ray region by high harmonic generation
(HHG). All major features of HHG, such as its comb-like spectrum of
odd-integer harmonics (to be called "IR harmonics"), its photons'
maximal energy (to be called "IR cut-off") and the emission times of
each harmonic, could be well reproduced using a semiclassical
three-step (recollision) model \cite{P. B. Corkum,M. Lewenstein,K.
J. Schafer}: under the influence of the intense laser field the
electron of an atom tunnels out of the modified Coulomb potential,
gains kinetic energy as a free particle in the field and finally may
recombine with the parent ion to release the sum of its kinetic
energy and the ionization potential as a high energy photon. The
emission times of different harmonics are perfectly synchronized
with the driver field, making the HHG process a promising method for
the production of an adjustable coherent X-ray source. The current
method of achieving the state of the art IR cut-off positions in the
harmonic generation spectra (HGS) makes use of high-intensity
few-femtosecond IR laser pulses. The main drawback of this method is
that the electronic plasma which is inevitably formed at such high
intensities, causes large dispersion on the propagating harmonics
and severely limits their phase matching. The method to achieve
higher-energy harmonics which will be presented here, doesn't suffer
from this limitation, since it allows the usage of an IR source of
moderate intensity which produces a small amount of plasma. It uses
an XUV driving field which is shined on the atom simultaneously with
the IR one. Taking the frequency of the XUV field larger than the
IR-cut off frequency and inside some spectral window of the HHG
generating gas, new higher order harmonics (to be called "XUV
harmonics"), which were absent in the spectra in the presence of the
IR field alone, could be produced, with frequencies well above the
IR cut-off frequency, while the XUV photoionization could be
suppressed, thus un-altering the amount of electronic plasma. As
will be shown, the main drawback of this method is that the new XUV
harmonics have a relatively low intensity.

The idea of contaminating the strong IR field with a second or more
higher-frequency [usually ultra-violet (UV)] fields, is not a new
one in the context of strong laser-matter interactions. The effect
on the dynamical behavior of the electrons is dramatic, and such
two-color (bichromatic) schemes have drawn a lot of attention in
recent years. The additional field, if adjusted correctly, can
induce stimulated emission \cite{Ph. Zeitoun}, single-photon
ionization \cite{Markus Kitzler}, or multi-photon ionization
\cite{N. A. Papadogiannis}. This ionization effect is actually
utilized for attosecond (as) pulse-duration measurement, by the
analysis of photoelectrons which are emitted from atoms exposed
simultaneously to the as-pulses and strong IR field
\cite{Hentschel}. The HGS obtained using bichromatic laser fields
\cite{T. Pfeifer,T. Pfeifer+L. Gallmann,N. Dudovich, H. Eichmann, M.
B. Gaarde+A. L'Huillier}, polychromatic fields \cite{Enrique
Conejero Jarque, M. B. Gaarde, C. Figueira}, or even as-pulses
\cite{Markus Kitzler} instead of the conventional monochromatic
field, had been studied extensively as well. To the best of our
knowledge, in all above-mentioned studies, the frequency of the UV
field was in the plateau of the HGS generated by the IR field alone
(to be called here "IR HGS"), or close to its cut-off \cite{C. Liu,
K. Ishikawa}. Such a UV field could only increase the efficiency of
the existing IR harmonics and/or create new peaks (hyper raman
lines) still in the support of the IR HGS, or maybe extend the IR
cut-off to some limited extent [As will be shown here, it is the
taking of the XUV field's frequency \textit{beyond} the IR-cut off
that pushes the cut-off position substantially to higher
frequencies]. On the basis of the three-step (re-collision) model,
it had been argued that the role of the UV field is to switch the
initial step in the generation of high harmonics from tunnel
ionization to the more efficient single UV-photon ionization
\cite{A. Heinrich}, or to assist the tunneling by transferring
population to an excited state, from which the tunneling rate is
larger \cite{A. Bandrauk, K. Ishikawa}, especially if the UV photon
energy matches some level transition \cite{K. Schiessl, K.
Ishikawa+K. Midorikawa}. This might explain the improved macroscopic
HHG signal obtained in experiments: the UV-assisted ionization
increases the number of atoms which participate in the HHG process
and improves phase matching (the possibility that the improvement in
HHG efficiencies is due to the interaction of the strong IR field
with the created ions was shown to be implausible \cite{C.
Valentin}). The above explanation doesn't apply, however, for the
case that will be discussed in this paper, i.e. a case in which the
high-frequency field in the bichromatic HHG scheme is in the XUV
regime (and not in the UV one), with frequency well above the IR
cut-off frequency. By choosing the frequency of the XUV field to
fall inside some spectral window of the HHG generating gas, the XUV
photoionization process could be eliminated (the same effect is
achieved automatically for high-enough energies of the XUV photon
since the single XUV photon ionization cross section scales as the
$7/2$-th power of the XUV photon's wavelength \cite{Sakurai}). Thus,
in this case, the contribution of the XUV field to the generation of
the XUV harmonics is not via affecting the ionization stage anymore.
Another suggestion for the role that the high-frequency photons play
in the HHG process was that they control the timing of ionization,
and preferentially select certain quantum paths of the electron
\cite{K. J. Schafer+M. B. Gaarde}. While this effect may lead to the
enhancement of the low-order harmonics in the plateau, it can't
account for the large enhancement in the cutoff and beyond (which is
noticeable in the results in Fig.\ref{fig1}).

A three-step model classical analysis of HHG suggests that the
contribution of the XUV field to the kinetic energy of the returning
electron is negligible. To see this, suppose we irradiate the atom
with a linearly-polarized IR fundamental field of frequency
$\omega$, amplitude $\varepsilon^{in}_{1}$ and polarization
$\mathbf{e_{k}}$
($\mathbf{E_{1}}(t)=\mathbf{e_{k}}\varepsilon^{in}_{1}\cos(\omega
t)$) and an XUV field of frequency $\tilde{q}\omega$ (where
$\tilde{q}$ is a large enough number) and amplitude
$\varepsilon^{in}_{\tilde{q}}$
($\mathbf{E_{\tilde{q}}}(t)=\mathbf{e_{k}}\varepsilon^{in}_{\tilde{q}}\cos(\tilde{q}\omega
t)$) with the same polarization. By integrating the classical
equation of motion while assuming that the electron is freed at time
$t_{i}$ with zero momentum, the following expression for the
momentum of the electron is obtained:

\begin{equation}
\mathbf{p}(t)=\mathbf{p_{1}}(t)+\mathbf{p_{\tilde{q}}}(t) \label{eq1}
\end{equation}
where $\mathbf{p_{1}}(t)$ is the momentum due to the IR field alone
$\mathbf{p_{1}}(t)=\frac{e\varepsilon^{in}_{1}}{\omega} [sin(\omega
t)-sin(\omega t_{i})]$ and $\mathbf{p_{\tilde{q}}}(t)$ is the
momentum due to the XUV field alone
$\mathbf{p_{\tilde{q}}}(t)=\frac{e\varepsilon^{in}_{\tilde{q}}}{\tilde{q}\omega}
[sin(\tilde{q}\omega t)-sin(\tilde{q}\omega t_{i})]$ ($e$ and $m$
are the electron's charge and mass, respectively). The kinetic
energy with two fields simultaneously present is:

\begin{equation}
E_{k}(t)=\frac{p^{2}(t)}{2m}=E_{k,1}(t)+E_{k,\tilde{q}}(t)+\frac{\mathbf{p_{1}}\cdot\mathbf{p_{\tilde{q}}}}{2m}
\label{eq2}
\end{equation}
where $E_{k,1}(t)=\frac{p_{1}^{2}(t)}{2m}$ and
$E_{k,\tilde{q}}(t)=\frac{p_{\tilde{q}}^{2}(t)}{2m}$. Note that
$E_{k,\tilde{q}}(t)\propto
(\frac{\varepsilon^{in}_{\tilde{q}}}{\tilde{q}\omega})^{2}$ and that
the cross-term
$\frac{\mathbf{p_{1}}\cdot\mathbf{p_{\tilde{q}}}}{2m}\propto
\frac{\varepsilon^{in}_{\tilde{q}}}{\tilde{q}\omega}$ are much
smaller than $E_{k,1}(t)$ if
$\frac{\varepsilon^{in}_{\tilde{q}}}{\tilde{q}}<\varepsilon^{in}_{1}$.
Usually, a large enough value of $\tilde{q}$ will fulfill this
condition, even if the IR and XUV fields have similar intensities.
Thus, the additional XUV field will not affect the electron
trajectories and will not contribute to their kinetic energy. For
this reason the relative phase between the two fields doesn't play a
role in the HGS, which is indeed verified in both classical analysis
and quantum mechanical simulations (a small $q$, however, will
affect the dynamics differently \cite{T. Pfeifer, Andiel}). In
addition, assigning the electron a non-zero initial momentum to
account for the photoelectric effect, will not increase its kinetic
energy upon recombination. To conclude, the extension of the
harmonic cutoff energy due to the inclusion of the XUV field
(Fig.\ref{fig1}), isn't a result of an increase of the electron's
kinetic energy upon recombination.

Since the XUV field doesn't affect the kinetic energy of the
electron trajectories (second step in the re-collision model), nor
does it modify the ionization step (first step in the re-collision
model), it must then influence the recombination step (third step in
the re-collision model) of HHG. It is the purpose of this paper to
prove this hypothesis. As will be shown later, the XUV field induces
periodic ac-Stark modulations to the remaining ground electronic
state, with the same frequency as the XUV field. The returning
electronic wavepacket recombines with this modulated ground state to
emit the new XUV harmonics. The IR-HGS enhancement and the cut-off
extension are a result of a single atom phenomenon, and not a
macroscopic one.

The paper is organized as follows: in section II we give the
numerical results of the HGS obtained using a model Hamiltonian
which describes a one-dimensional Xe atom subjected to a sine-square
pulse of bichromatic field of frequencies $\omega$ and
$\tilde{q}\omega$, for different values of $\tilde{q}$. In section
III we briefly describe the semiclassical re-collision model of HHG,
as is usually applied to the monochromatic case. We then modify the
re-collision model to account for possible non-trivial
time-dependence of the ground electronic state, and show that this
modified re-collision model successfully reproduces the results
presented in section II. In section IV we suggest an experiment
based on the effect we discovered and conclude.

\section{Xe atom driven by a two-color $(\omega,\tilde{q}\omega)$
laser field}

As an illustrative numerical demonstration of the IR cut-off
extension in the HGS due to the addition of an XUV field, we studied
a single electron 1D Xe atom irradiated by a sine-square pulse
supporting $N$ oscillations of linearly polarized light of
bichromatic field, composed of an IR laser field of frequency
$\omega$ and amplitude $\varepsilon^{in}_{1}$ and a high-frequency
field of frequency $\tilde{q}\omega$ and amplitude
$\varepsilon^{in}_{\tilde{q}}$. The following time-dependent
Schr\"{o}dinger equation (TDSE) was integrated using the split
operator method:

\begin{equation}
i\hbar\frac{\partial}{\partial
t}\Psi(x,t)=\biggl\{\frac{p_{x}^{2}}{2m}+V_{0}(x)-ex
sin^{2}\biggl(\frac{\omega t}{2N}\biggr) [\varepsilon^{in}_{1}
\cos(\omega t)+\varepsilon^{in}_{\tilde{q}} \cos(\tilde{q}\omega
t)]\biggr\}\Psi(x,t) \label{eq3}
\end{equation}
between the times $0<t<NT$ ($T=\frac{2\pi}{\omega}$) with atomic
units ($\hbar=m=-e=1$) and with the wave function taken initially as
the ground state $\phi_{1}(x)$ of the field-free model Hamiltonian
of a 1D Xe atom, with the field-free effective potential
$V_{0}(x)=-0.63exp(-0.1424x^{2})$. This potential supports three
bound states, of which the two lowest ones mimic the two lowest
electronic states of Xe, with energies $-I_{p}=-0.4451 a.u.$ and
$E_{2}=-0.1400 a.u.$. The parameters used for the simulation were
$N=50$, $\omega=0.05695a.u.$ ($\lambda =800nm$),
$\varepsilon^{in}_{1}=0.035a.u.$ (corresponding to intensity of
$I^{in}_{1}\simeq 4.299\cdot10^{13}\frac{W}{cm^{2}}$),
$\varepsilon^{in}_{\tilde{q}}=0.0001a.u.$ ($I^{in}_{\tilde{q}}\simeq
3.509\cdot 10^{8}W/cm^{2}$). It should be noted that in this type of
simulation the Born approximation is assumed: instead of solving two
coupled differential equations, one for the evolution of the
electron and the other for the propagation of the electromagnetic
field, the electromagnetic field is assumed to remain unchanged
during the interaction with the electron, and only a Schr\"{o}dinger
equation of motion for the electron is solved.

In order to calculate the HGS the Larmor approximation
\cite{Jackson} was assumed, and the time-dependent acceleration
expectation value

\begin{equation}
a(t)\equiv\frac{1}{m}\langle \Psi(x,t)|-
\frac{dV_{0}(x)}{dx}|\Psi(x,t) \rangle
+\frac{e}{m}sin^{2}\biggl(\frac{\omega t}{2N}\biggr) [\varepsilon^{in}_{1}
\cos(\omega t)+\varepsilon^{in}_{\tilde{q}} \cos(\tilde{q}\omega
t)]\label{eq4}
\end{equation}
which is linearly proportional to the emitted field, was analyzed.
The power spectra (HGS) of emitted radiation by the oscillating
electron is proportional to the modulus-square of the
Fourier-transformed time-dependent acceleration expectation value:

\begin{equation}
\sigma(\Omega)=\frac{2e^{2}}{3c^{3}} |a(\Omega)|^{2}
\label{eq5}
\end{equation}
where the acceleration in frequency space is given by the Fourier
transform

\begin{equation}
a(\Omega)=\frac{1}{NT}\int_{0}^{NT}
a(t)e^{-i\Omega t } dt \label{eq6}
\end{equation}

\begin{figure}[ht]
\hbox{\includegraphics[width=3.8in,height=3.8in]{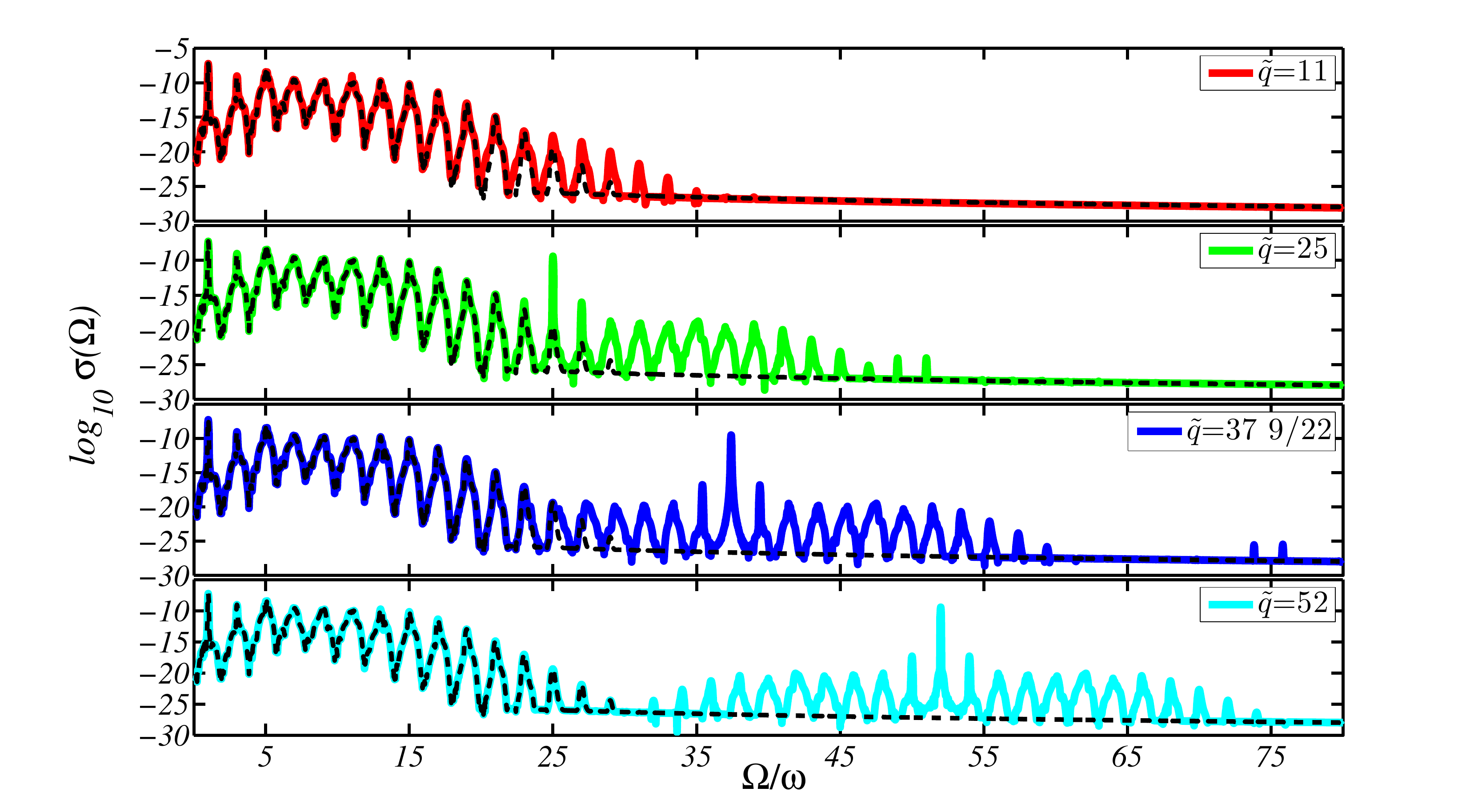}}%[width=3.8in,height=2.8in]
\caption{\label{fig1}\small (color online) HGS obtained from a 1D
model Hamiltonian of Xe atom (Eq.\ref{eq3}) irradiated by a 50-oscillation
sine-square pulse of bichromatic laser field composed of a 800nm
IR laser field of intensity $I^{in}_{1}\simeq 4.299\cdot 10^{13}W/cm^{2}$ and a $800/\tilde{q}$-nm XUV
field of intensity $I^{in}_{\tilde{q}}\simeq 3.509\cdot 10^{8}W/cm^{2}$ for different values
of $\tilde{q}$: $\tilde{q}=11$ (solid red line), $\tilde{q}=25$ (solid green line),
$\tilde{q}=37\frac{9}{22}$ (solid blue line) and $\tilde{q}=52$ (solid cyan line). The IR-HGS is
shown as the dotted black line where the position of the IR cutoff is
at the 15th harmonic. The addition of the XUV field could have several effects on the HGS,
depending on its frequency with respect to the position of the IR cut-off: it could
either modify the IR-HGS or slightly extend the IR cut-off ($\tilde{q}=11$), or add new XUV harmonics to the spectrum ($\tilde{q}=25, 37\frac{9}{22}, 52$).
The XUV harmonics appear at $\tilde{q} \pm 2K$, regardless of the value of $\tilde{q}$. The XUV-HGS possesses the following symmetries:
with respect to its center $\tilde{q}$, the distribution of the new XUV harmonics is
symmetric (i.e., for $\tilde{q}=52$,
$\sigma(48\omega)\simeq\sigma(56\omega)$, etc.) and upon variation
of $\tilde{q}$ it shifts but remains almost invariant. }
\end{figure}

Fig.\ref{fig1} shows the HGS for different values of $\tilde{q}$
($\tilde{q}=11$, $\tilde{q}=25$, $\tilde{q}=37\frac{9}{22}$ and
$\tilde{q}=52$). The HGS in the presence of the IR field alone is
also shown for comparison. Several features can be seen in the
figure: The position of the IR cut-off (the cut-off of the HGS in
the presence of the IR field only) is at the 15th harmonic where the
maximal IR harmonic is the 29th one. When the high frequency field
has a frequency still within the IR-HGS, its only influence on the
HGS is to modify the IR-harmonics ($\tilde{q}=11$ in
Fig.\ref{fig1}). When the high frequency field has a frequency close
to the IR cut-off, the IR-harmonics are modified, and new harmonics
(above the 29th harmonic), which were not present with the IR field
alone, appear ($\tilde{q}=25$ in Fig.\ref{fig1}, which corresponds
to an XUV radiation of wavelength 16nm). These new harmonics, that
appear due to the addition of the XUV field only, will be termed
XUV-harmonics. The position of the maximal XUV harmonic increases as
the XUV frequency increases ($\tilde{q}=37\frac{9}{22}$), where it
becomes apparent that the XUV harmonics appear around the frequency
of the XUV field. The number of new XUV harmonics reaches a maximal
one (approximately twice the number of IR harmonics) whenever the
value of $\tilde{q}$ is either not odd integer or is greater than
approximately twice the number of the maximal IR harmonic
($\tilde{q}=52$ in Fig.\ref{fig1}).

In general, upon the addition of the XUV field of frequency
$\tilde{q}\omega$ to the IR field of frequency $\omega$ the
harmonics $\tilde{q} \pm 2K$ ($K$ integer) are either modified (if
they were already present in the IR-HGS) and/or appear as new XUV
harmonics. Moreover, the HGS possesses certain symmetries: with
respect to its center at harmonic $\tilde{q}$, the distribution of
the XUV harmonics is symmetric (i.e., for
$\tilde{q}=37\frac{9}{22}$,
$\sigma(33\frac{9}{22}\omega)\simeq\sigma(41\frac{9}{22}\omega)$,
etc.) and upon variation of $\tilde{q}$ it shifts but remains almost
invariant. The structure of the HGS of the XUV harmonics (will be
called XUV-HGS from now on) consists, in principal, of two new
plateau-like regions and two new cut-off-like regions. For example,
for $\tilde{q}=52$ in Fig.\ref{fig1}, the harmonics of order 38-48
and 56-66 have a "plateau" character (constant intensity), and the
harmonics 32-36 and 68-72 have a "cut-off" character (constant
phase, will be shown in Fig.\ref{fig5}). The XUV harmonics are
10-orders of magnitude weaker than the IR harmonics.

\begin{figure}[ht]
\hbox{\includegraphics[width=3.8in,height=2.8in]{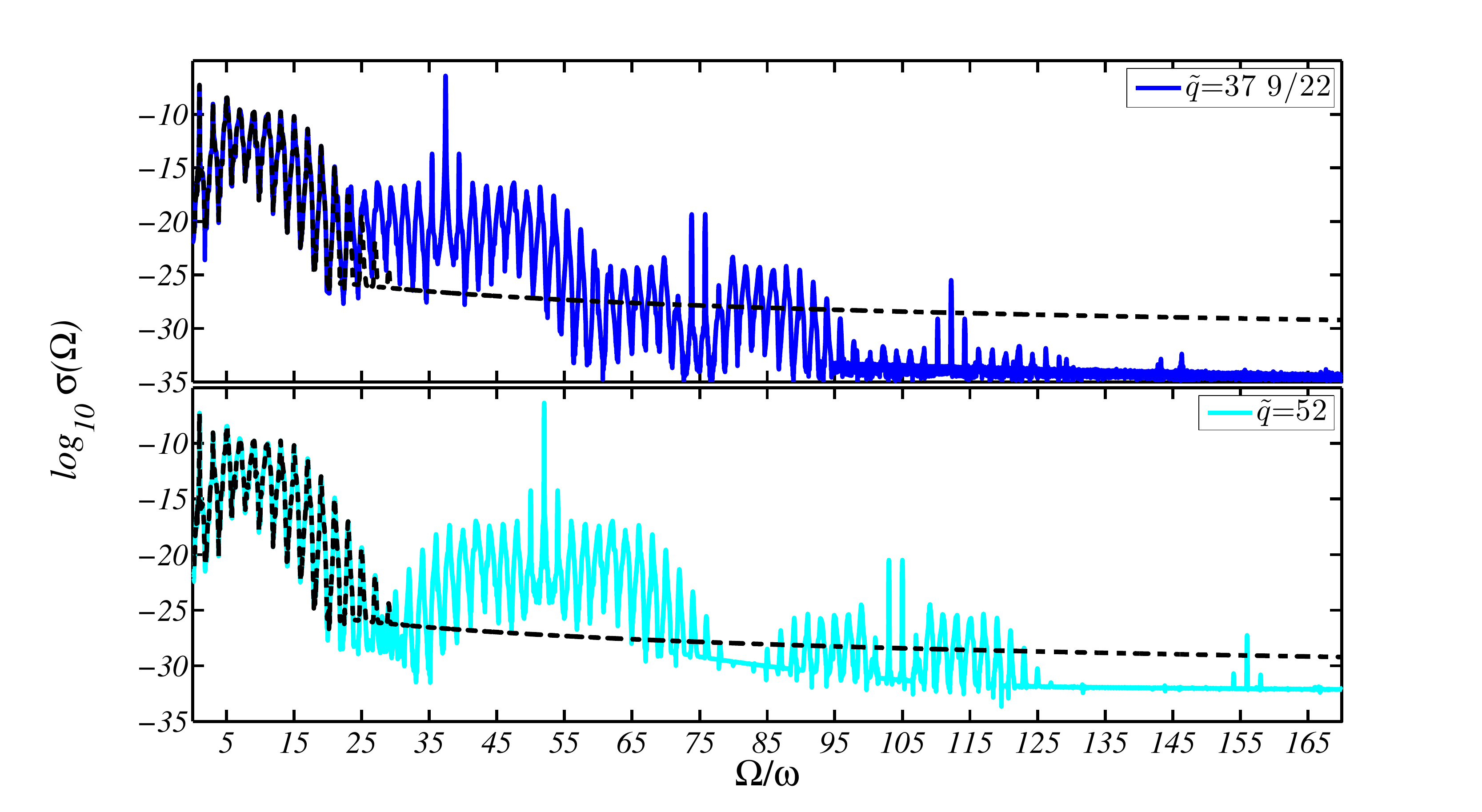}}
\caption{\label{fig2}\small (color online) Same as in Fig.\ref{fig1}, but for $I^{in}_{\tilde{q}}\simeq 4.299\cdot 10^{11}W/cm^{2}$.
For $\tilde{q}=37\frac{9}{22}$ (solid blue line) three sets of XUV harmonics are observed: one at $\tilde{q} \pm 2K$, a second one at $2\tilde{q} \pm (2K-1)$, and a third one at
$3\tilde{q} \pm 2K$, all differ very much in their intensities but have the same general structure and the same
symmetries pointed out in Fig.\ref{fig1} and in the text. For $\tilde{q}=52$ (solid cyan line) only 2 sets of
XUV harmonics are observed, together with traces of the third set. The constant intensity of the IR-HGS (dotted black line)
above the IR cut-off should be disregarded as it is due to numerical error and could be made as low as one wishes.}
\end{figure}

Fig.\ref{fig2} shows the HGS for a larger intensity of the
high-frequency field, $\varepsilon^{in}_{\tilde{q}}=0.0035a.u.$
($I^{in}_{\tilde{q}}\simeq 4.299\cdot 10^{11}W/cm^{2}$), where all
other parameters are kept the same. For sake of clarity, only two
values of $\tilde{q}$ are shown $\tilde{q}=37\frac{9}{22}$ and
$\tilde{q}=52$. As before, the set of new XUV harmonics $\tilde{q}
\pm 2K$ ($K$ integer) appears around $\tilde{q}$. It is 7-orders of
magnitude weaker than the IR harmonics. In addition, an additional
set of new XUV harmonics $2\tilde{q} \pm (2K-1)$ ($K$ integer)
appears around $2\tilde{q}$ and possesses the same above-mentioned
symmetries: it consists of two new plateau-like regions and two
cut-off-like regions. The intensity of the XUV harmonics around
$2\tilde{q}$ are, however, 14-orders of magnitude weaker than the IR
harmonics. A third set of XUV harmonics $3\tilde{q} \pm 2K$ appears
around $3\tilde{q}$. It possesses the same above-mentioned
symmetries and is 21-orders of magnitude weaker than the IR
harmonics.

\begin{figure}[ht]
\hbox{\includegraphics[width=3.8in,height=2.8in]{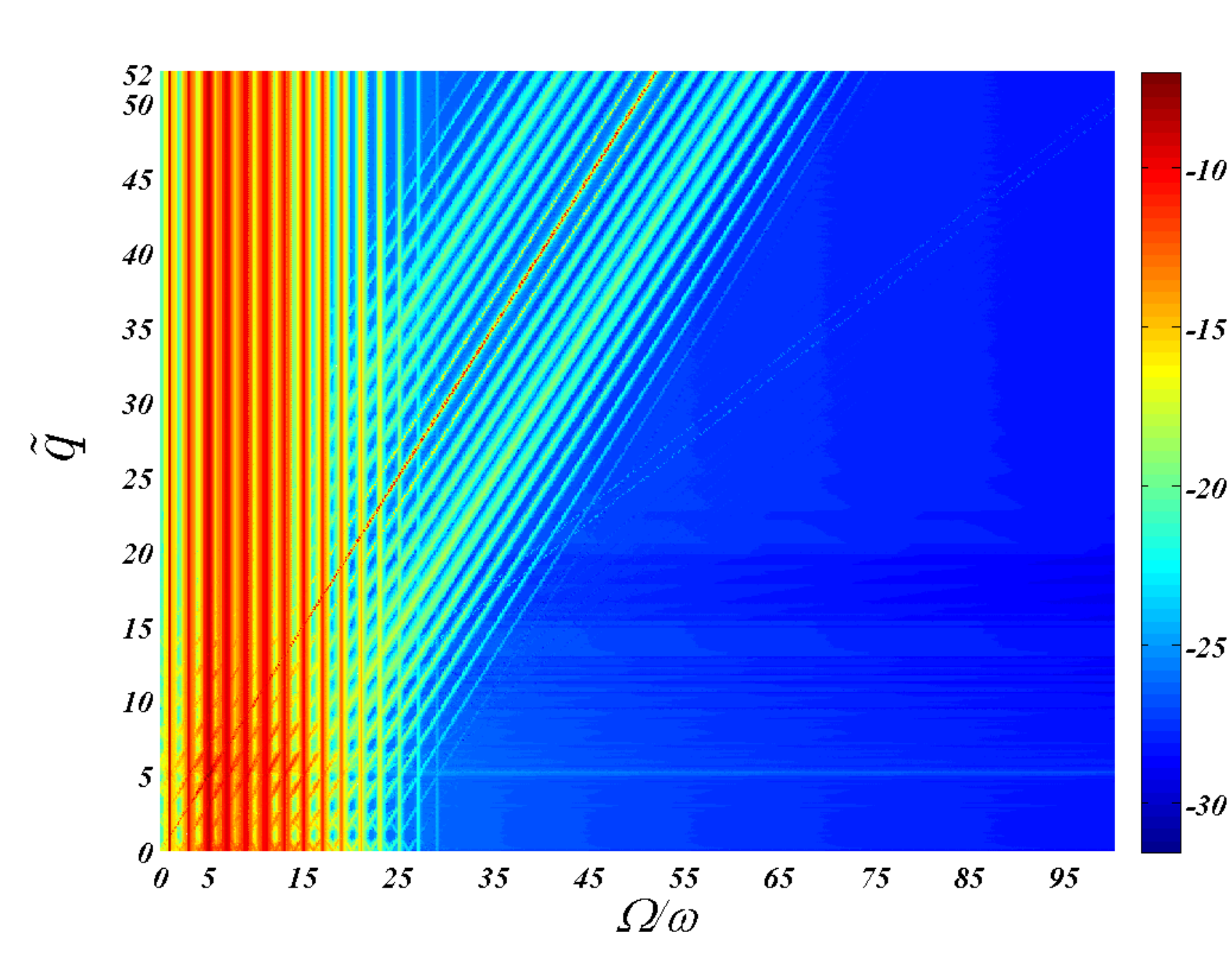}}
\caption{\label{fig3}\small (color online) Top view of a log plot showing the
HGS (red color-high intensity, blue color-low intensity) obtained from a 1D
model Hamiltonian of Xe atom (Eq.\ref{eq3}) irradiated by a 50-oscillation
sine-square pulse of bichromatic laser field composed of a 800nm
IR laser field of intensity $I^{in}_{1}\simeq 4.299\cdot 10^{13}W/cm^{2}$ and a $800/\tilde{q}$nm XUV
field of intensity $I^{in}_{\tilde{q}}\simeq 3.509\cdot 10^{8}W/cm^{2}$ for different values
of $\tilde{q}$. The HGS shown in Fig.\ref{fig1} is obtained by taking cuts of the HGS shown here along the specific values of $\tilde{q}$.
The "lines" in the spectrum are merely a manifestation of the selection rules given in Eq.\ref{eq7}.
Groups of harmonics corresponding to absorption of more than one XUV photon are absent, due to the
weak intensity of the XUV field. That is, the only set of XUV harmonics that appear is the set $\tilde{q} \pm 2K$. }
\end{figure}

\begin{figure}[ht]
\hbox{\includegraphics[width=3.8in,height=2.8in]{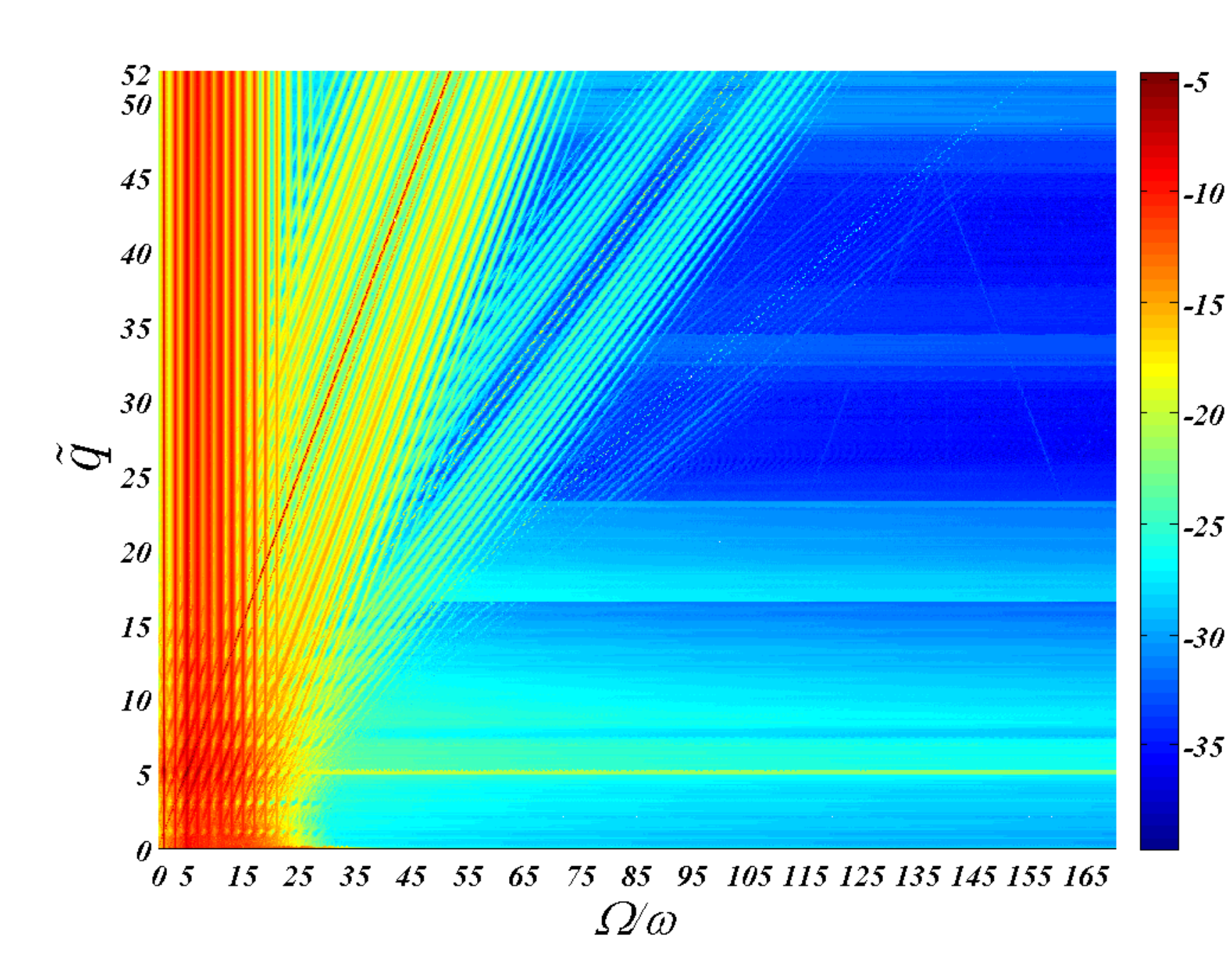}}
\caption{\label{fig4}\small (color online) The same as in Fig.\ref{fig3} but with a larger intensity of the second field:
$I^{in}_{\tilde{q}}\simeq 4.299\cdot 10^{11}W/cm^{2}$. This time, also the sets of XUV harmonics $2\tilde{q} \pm (2K-1)$ and $3\tilde{q} \pm 2K$ appear. }
\end{figure}

For a larger intensity of the high frequency field this same trend
is continued, giving rise to new sets of XUV harmonics around
$4\tilde{q} \pm (2K-1)$, $5\tilde{q} \pm 2K$, etc. which could
become nested but are well distinguished by the large differences in
their intensities. Fig.\ref{fig3} and Fig.\ref{fig4}, which show a
top-view plot of the bichromatic HGS as function of $\tilde{q}$,
suggest that the new sets of XUV harmonics emerge from the single
set of IR harmonics. The sets of straight lines with different
slopes in those plots correspond to different values of the integers
$n_{1}$ and $n_{\tilde{q}}$ in the selection rules applicable for
our bichromatic $(\omega,\tilde{q}\omega)$ scheme \cite{Avner Atto}:
the possible harmonics that could be emitted are
$\Omega/\omega=n_{1}+n_{\tilde{q}}\tilde{q}$ (where
$n_{1}+n_{\tilde{q}}=2K-1$), or alternatively

\begin{equation}
\tilde{q}=\frac{1}{n_{\tilde{q}}}\frac{\Omega}{\omega}-\frac{n_{1}}{n_{\tilde{q}}}
\label{eq7}
\end{equation}
We shall symbolize the sets of lines according to their values of
$n_{1}$ and $n_{\tilde{q}}$ as $(n_{1},n_{\tilde{q}})$. The set of
lines parallel to the vertical axis of Fig.\ref{fig3} (infinite
slope) correspond to $n_{\tilde{q}}=0$ and odd values of $n_{1}$,
i.e. to the IR harmonics $\Omega=(2K-1)\omega$ (in a perturbative
picture, this corresponds to no absorption of $\tilde{q}\omega$
photons, only an odd number of $\omega$ photons) and will be
symbolized as $(n_{1},n_{\tilde{q}})=(2K-1,0)$. It is obvious that
without the XUV field, only this set would appear in this type of
figure. The next set of lines, with slope equal to unity, correspond
to $(n_{1},n_{\tilde{q}})=(2K,1)$, i.e. to the XUV harmonics
$\Omega=(\tilde{q} \pm 2K)\omega$ that are shown also in
Fig.\ref{fig1}. It is seen that these lines are much weaker than the
lines corresponding to the set $(n_{1},n_{\tilde{q}})=(2K-1,0)$. The
strongest line in the set $n_{\tilde{q}}=1$ corresponds to
$(n_{1},n_{\tilde{q}})=(0,1)$, i.e. absorption of one
$\tilde{q}\omega$ photon and no $\omega$ photons. This line,
together with the lines to its right [which correspond to
$(n_{1},n_{\tilde{q}})=(2,1), (4,1),...$] emerge from the same
points on the function $\tilde{q}=1$ form where the IR harmonic
lines emerge. The lines $(n_{1},n_{\tilde{q}})=(-2,1), (-4,1),...$
are a mirror image of the lines
$(n_{1},n_{\tilde{q}})=(2,1),(4,1),...$, with respect to the central
line $(n_{1},n_{\tilde{q}})=(0,1)$. This hints us that the symmetry
properties of the XUV-HGS (i.e., the fact that the distribution of
the XUV harmonics is symmetric with respect to $\tilde{q}$, shifts
but remains almost invariant upon variation of $\tilde{q}$, and
consists of two "plateau" and two "cut-off" regions) are the result
of the fact that the XUV harmonics are "born" from the same
electronic trajectories that produce the IR harmonics. When the
intensity of the high-frequency field is increased, as in
Fig.\ref{fig4}, also the sets $(n_{1},n_{\tilde{q}})=(2K-1,2)$ and
$(n_{1},n_{\tilde{q}})=(2K,3)$ appear, and they are 20-orders and
30-orders of magnitude smaller than the IR harmonics, as shown in
Fig.\ref{fig2}. Also these XUV harmonics are "born" from the
electronic trajectories that produce the IR harmonics. What is the
physical mechanism leading to the formation of these new sets of XUV
harmonics?

\section{The re-collision description of bichromatic $(\omega,\tilde{q}\omega)$ HHG}

The process of HGS could be successfully described in terms of a
very simple and intuitive model: the semiclassical re-collision
model \cite{P. B. Corkum}. Let us describe, for the beginning, the
process of HHG driven by an IR filed only. According to this model,
the electronic wavefunction at the event of recombination
$\Psi(\mathbf{r},t\approx t_{r})$ could be described as a sum of the
following continuum and bound parts

\begin{equation}
\Psi(\mathbf{r},t_{r})=\psi_{b}(\mathbf{r},t_{r})+\psi_{c}(\mathbf{r},t_{r})
\label{eq8}
\end{equation}

It is assumed that the strong IR field ionizes the electron by
tunneling from the initial ground state of the field-free
Hamiltonian $\phi_{1}(\mathbf{r})$, which is only slightly depleted
during this process. It is assumed that the electronic wavefunction
which remain bound, evolves under the field-free Hamiltonian only,
i.e. accumulates a trivial phase only:

\begin{equation}
\psi_{b}(\mathbf{r},t)=\phi_{1}(\mathbf{r})e^{\frac{i}{\hbar}I_{p}t}
\label{eq9}
\end{equation}
where $-I_{p}$ is the energy of the ground state. Under the strong
field approximation, the freed electronic continuum part evolves
under the external field only. Taking the direction of linear
polarization $\mathbf{e_{k}}$ as the x-direction from now on for
simplicity, and assuming separability of the continuum wavefunction
$\psi_{c}(\mathbf{r},t\approx t_{r})$ in the x-coordinate and the 2
other lateral coordinates for simplicity, the continuum wavefunction
can be written as

\begin{equation}
\psi_{c}(\mathbf{r},t\approx t_{r})=\psi_{c}^{\parallel}(x,t\approx t_{r})\psi_{c}^{\perp}(y,z,t\approx t_{r})
\label{eq10}
\end{equation}
where the returning continuum part in the direction of polarization
is some superposition of plane waves

\begin{equation}
\psi_{c}^{\parallel}(x,t_{r})=\frac{1}{\sqrt{2\pi}}\int_{-\infty}^{\infty}dk
\tilde{\psi_{c}}^{\parallel}(k,t_{r})e^{i[kx-\frac{E_{k}}{\hbar} t_{r}]}
\label{eq11}
\end{equation}
where $\mathbf{k}=k\mathbf{e_{x}}$ ($k=|\mathbf{k}|$) is the
momentum of the electron, $E_{k}\equiv \frac{\hbar^{2}k^{2}}{2m}$ is
the usual dispersion relation and
$\tilde{\psi_{c}}^{\parallel}(k,t_{r})$ are expansion coefficients
which weakly depend on time.

Using the total wavefunction at the event of recombination
$\Psi(\mathbf{r},t\approx t_{r})$, the time-dependent acceleration
expectation value could be calculated. Keeping only the part which
is responsible for the emission of radiation at frequencies other
than the incident frequency $\omega$, the acceleration reads
$\mathbf{a}(t)\equiv\frac{1}{m}\langle \Psi(\mathbf{r},t)|-\nabla
V_{0}(\mathbf{r})|\Psi(\mathbf{r},t) \rangle_{\mathbf{r}}$, where
$V_{0}(\mathbf{r})$ is the field-free potential. Assuming low
depletion rate of the ground state (and hence, small population of
the continuum wavepacket), the dominant terms that are responsible
for the emission of radiation at frequencies other than the incident
frequency $\omega$ are the bound-continuum terms

\begin{equation}
\mathbf{a}(t_{r})= \frac{2}{m}\Re \langle \psi_{b}(\mathbf{r},t_{r})|-\nabla
V_{0}(\mathbf{r})|\psi_{c}(\mathbf{r},t_{r}) \rangle_{\mathbf{r}}
\label{eq12}
\end{equation}
(the bound-bound term $\langle \psi_{b}(\mathbf{r},t)|-\nabla
V_{0}(\mathbf{r})|\psi_{b}(\mathbf{r},t) \rangle_{\mathbf{r}}$ is
time-independent and doesn't radiate and the contribution of the
continuum-continuum term is negligible). After plugging the
expressions in Eq.\ref{eq9}-\ref{eq11} into Eq.\ref{eq12}, it could
be realized that the acceleration is composed of oscillating terms
of the form

\begin{equation}
\mathbf{a}(t_{r})= \Re \int_{-\infty}^{\infty}dk \tilde{\tilde{\psi}}^{(0)}(k,t_{r})
e^{-\frac{i}{\hbar}[I_{p}+\frac{\hbar^{2}k^{2}}{2m}]t_{r}} \label{eq13}
\end{equation}
where

\begin{equation}
\tilde{\tilde{\psi}}^{(0)}(k,t_{r})\equiv -\frac{2}{m}\frac{1}{\sqrt{2\pi}}
\tilde{\psi_{c}}^{\parallel}(k,t_{r})\int_{-\infty}^{\infty} d^{3}r
\phi_{1}(\mathbf{r})\nabla V_{0}(\mathbf{r})
\psi_{c}^{\perp}(y,z,t_{r})e^{ikx} \label{eq14}
\end{equation}

The emitted HHG field in a single re-collision event at $t\approx
t_{r}$ is a burst of light which corresponds to the spectral
continuum $I_{p}<\hbar\Omega<I_{p}+3.17U_{p}$ where $3.17U_{p}\equiv
3.17\frac{e^{2}(\varepsilon^{in}_{1})^{2}}{4m \omega^{2}}$ is the
value of the most energetic returning electron trajectory. In
addition, since we assumed that the IR field pulls the electron
along the x-direction, it is reasonable to assume symmetric
evolution of the continuum wavefunction in the lateral plane, i.e.,
$\psi_{c}^{\perp}(y,z,t_{r})=\psi_{c}^{\perp}(-y,-z,t_{r})$. Since
for atoms $V_{0}(\mathbf{r})$ and $\phi_{1}(\mathbf{r})$ are
symmetric functions (and $\nabla V_{0}(\mathbf{r})$ is
antisymmetric), we get from Eq.\ref{eq14} that the coefficient
$\tilde{\tilde{\psi}}^{(0)}(k,t_{r})$ has a nonzero component along
the x-direction only, i.e., the acceleration $\mathbf{a}(t_{r})$
points along the x-direction, as it should.

When each single re-collision event is repeated every half cycle of
the IR field, integer odd harmonics $2K-1$ are obtained in the HGS.
To see this we compare two consecutive re-collision events at times
$t_{r}$ and $t_{r}+\frac{T}{2}$. Suppose we assume that
$\psi_{c}^{\parallel}(x,t_{r})$ was born at some initial time
$t_{i}$ from $\phi_{1}(\mathbf{r})$. Therefore, since
$\psi_{c}^{\parallel}(x,t_{r}+\frac{T}{2})$ was born $\frac{T}{2}$
after $\psi_{c}^{\parallel}(x,t_{r})$, at the time
$t_{i}+\frac{T}{2}$, it was born from
$\phi_{1}(\mathbf{r})e^{\frac{i}{\hbar}I_{p}\frac{T}{2}}$ since the
bound state from which the continuum state tunnels out, has
accumulated this phase. In addition, the two continuum functions
$\psi_{c}^{\parallel}(x,t_{r})$ and
$\psi_{c}^{\parallel}(x,t_{r}+\frac{T}{2})$ are released in
\textit{opposite} spatial directions, because the IR field changes
direction in two subsequent tunneling times. Therefore, the symmetry
relation between $\psi_{c}^{\parallel}(x,t_{r})$ and
$\psi_{c}^{\parallel}(x,t_{r}+\frac{T}{2})$ is

\begin{equation}
\psi_{c}^{\parallel}(x,t_{r})=\psi_{c}^{\parallel}(-x,t_{r}+\frac{T}{2})e^{-\frac{i}{\hbar}I_{p}\frac{T}{2}}
\label{eq15}
\end{equation}

Plugging this symmetry into Eq.\ref{eq11}, and using the property
$E_{k}\equiv \frac{\hbar^{2}k^{2}}{2m}= E_{-k}$, we get by simple
change of variables of the integral that

\begin{equation}
\tilde{\psi_{c}}^{\parallel}(k,t_{r})=\tilde{\psi_{c}}^{\parallel}(-k,t_{r}+T/2)e^{-\frac{i}{\hbar}(I_{p}+E_{k})\frac{T}{2}}
\label{eq16}
\end{equation}

Using the relations mentioned before
($V_{0}(\mathbf{r})=V_{0}(-\mathbf{r})$,
$\phi_{1}(\mathbf{r})=\phi_{1}(-\mathbf{r})$, $\nabla
V_{0}(\mathbf{r})=-\nabla V_{0}(-\mathbf{r})$,
$\psi_{c}^{\perp}(y,z,t_{r})=\psi_{c}^{\perp}(y,z,t_{r}+T/2)=\psi_{c}^{\perp}(-y,-z,t_{r})$)
the following symmetry is obtained from Eq.\ref{eq14}:

\begin{equation}
\tilde{\tilde{\psi}}^{(0)}(k,t_{r})=-\tilde{\tilde{\psi}}^{(0)}(-k,t_{r}+T/2)e^{-\frac{i}{\hbar}(I_{p}+E_{k})\frac{T}{2}}
\label{eq17}
\end{equation}

Using this symmetry, we get from Eq.\ref{eq13} that:

\begin{equation}
\mathbf{a}(t_{r}+T/2)= -\mathbf{a}(t_{r}) \label{eq18}
\end{equation}
The acceleration vector is periodic in $T$ and alternates directions
between subsequent re-collision events, i.e., its only nonzero
components in its Fourier expansion correspond to odd integer
harmonics of $\omega$. This is the origin of the well-known odd
selection rules of monochromatic HHG. We shall therefore symbolize
the acceleration which is responsible for the emission of odd
harmonics (Eq.\ref{eq13}) as $\mathbf{a}^{(0)}(t)$.

The three-step model described above assumes that the only time
evolution of the remaining bound part of the electronic wavefunction
is to accumulate a trivial phase, as given in Eq.\ref{eq9}. This
assumption, however, describes only the leading term in the time
evolution of the bound part. In reality, due to the ac-Stark effect
induced by the IR field, the electron adiabatically follows the
instantaneous ground state of the potential which periodically
shakes back and forth by the IR field. If one carries out a TDSE
simulation and looks at the electronic wavefunction in the
field-free potential region during the action of the IR field, one
sees that it oscillates back and forth with the same frequency of
the IR field. The time evolution of the bound part
$\psi_{b}(\mathbf{r},t)$ should be therefore corrected from the
trivial one given in Eq.\ref{eq9}. In addition, for common field
intensities, the ac-Stark correction to the instantaneous ground
state energy is negligible, and we may therefore assume that the
instantaneous ground-state energy is almost constant ($I_{p}$). More
importantly, the ac-Stark effect induces a periodic motion of the
wavefunction as a whole, without deforming it. Relying on these
facts, we approximate the instantaneous ground state wavefunction as

\begin{equation}
\psi_{b}(\mathbf{r},t)\cong \phi_{1}(x+\varepsilon_{1}^{out}
\cos(\omega t),y,z)
e^{+\frac{i}{\hbar}I_{p}t} \label{eq19}
\end{equation}

We have assumed the simplest time-dependence in
$\psi_{b}(\mathbf{r},t)$ that would still give periodic modulations
at frequency $\omega$. It should be noted that in the language of
non-Hermitian quantum mechanics \cite{Reviewnimrod, Avner Rapid}
this expression approximately describes the resonance Floquet state
which evolves from the ground state $\phi_{1}(\mathbf{r})$ upon the
switching of the IR field.

The quiver amplitude $\varepsilon_{1}^{out}$ of the spatial
oscillations of the ground state is of the order of
$\varepsilon_{1}^{out}=\frac{\varepsilon_{1}^{in}}{(E_{2}+I_{p})^{2}-\omega^{2}}$
(this is approximately the quiver amplitude of a an electron bound
in a short-range potential of the type used here, driven by an IR
field of amplitude $\varepsilon_{1}^{in}$), i.e., a tiny fraction of
a Bohr radius, provided that the laser's frequency doesn't match
some level transition. The bound part may therefore be expanded in a
Taylor serie as

\begin{equation}
\psi_{b}(\mathbf{r},t)\cong e^{+\frac{i}{\hbar}I_{p}t}
\{\phi_{1}(\mathbf{r})+\varepsilon_{1}^{out} \cos(\omega
t)\frac{\partial}{\partial
x}\phi_{1}(\mathbf{r})\} \label{eq20}
\end{equation}

Calculation of the time-dependent acceleration expectation value
using the total wavefunction at the event of recombination
$\Psi(\mathbf{r},t\approx t_{r})$ with the modified bound part
$\psi_{b}(\mathbf{r},t_{r})$ given in Eq.\ref{eq20}, and keeping
again only the terms that are responsible for the emission of
radiation at frequencies other than the incident frequency $\omega$
(the bound-continuum terms), yields:

\begin{equation}
\mathbf{a}(t_{r})= \mathbf{a}^{(0)}(t_{r})+\varepsilon_{1}^{out}\cos(\omega t_{r})\mathbf{a}^{(1)}(t_{r})
\label{eq21}
\end{equation}
where $\mathbf{a}^{(0)}(t_{r})$ is given in Eq.\ref{eq13} and

\begin{equation}
\mathbf{a}^{(1)}(t_{r}) = \Re \int_{-\infty}^{\infty} dk \tilde{\tilde{\psi}}^{(1)}(k,t_{r}) e^{-\frac{i}{\hbar}[I_{p}+\frac{\hbar^{2}k^{2}}{2m}]t_{r}}
\label{eq22}
\end{equation}
where

\begin{equation}
\tilde{\tilde{\psi}}^{(1)}(k,t_{r})\equiv -\frac{2}{m}\frac{1}{\sqrt{2\pi}}
\tilde{\psi_{c}}^{\parallel}(k,t_{r})\int_{-\infty}^{\infty} d^{3}r
\frac{\partial \phi_{1}(\mathbf{r})}{\partial x}\nabla V_{0}(\mathbf{r})
\psi_{c}^{\perp}(y,z,t_{r})e^{ikx} \label{eq23}
\end{equation}

We see that the inclusion of the Stark effect contributes a new term
$\varepsilon_{1}^{out}\cos(\omega t_{r})\mathbf{a}^{(1)}(t_{r})$ to
the acceleration. In a single re-collision event at $t\approx
t_{r}$, this term produces two bursts of light. One corresponds to
the spectral continuum
$I_{p}+\hbar\omega<\hbar\Omega<I_{p}+3.17U_{p}+\hbar\omega$ and the
other to
$I_{p}-\hbar\omega<\hbar\Omega<I_{p}+3.17U_{p}-\hbar\omega$. These
bursts of light are, however, much weaker than the one which results
from $\mathbf{a}^{(0)}(t_{r})$, since, as we recall, the factor
$\varepsilon_{1}^{out}=\frac{\varepsilon_{1}^{in}}{(E_{2}+I_{p})^{2}-\omega^{2}}$
is small for common IR laser frequency and intensity (in the context
of HHG experiments). It can be said in general that each electron
trajectory (plane wave) with kinetic energy $E_{k}$, recombines with
the nucleus to emit radiation of energy $I_{p}+E_{k}$, and two
"duplicate" photons with energies $I_{p}+E_{k}-\hbar\omega$ and
$I_{p}+E_{k}+\hbar\omega$, at the same emission times. In a multi
recollision sequence, only some of these photons will appear in the
HGS, as dictated by selection rules which we are about to prove. In
addition, since the functions $\frac{\partial
\phi_{1}(\mathbf{r})}{\partial x}$ and $\psi_{c}^{\perp}(y,z,t_{r})$
are symmetric with respect to $y,z$ and $\nabla
V_{0}(\mathbf{r})=-\nabla V_{0}(-\mathbf{r})$, we get from
Eq.\ref{eq22}-\ref{eq23} that the coefficient
$\tilde{\tilde{\psi}}^{(1)}(k,t_{r})$ and the acceleration
$\mathbf{a}^{1}(t_{r})$ point along the x-direction, as they should.

When each single re-collision event is repeated every half cycle of
the IR field, the new term in the acceleration contributes integer
odd harmonics $2K-1$ to the HGS. To see this we compare two
consecutive re-collision events at times $t_{r}$ and
$t_{r}+\frac{T}{2}$ and note that the symmetry relation between
$\psi_{c}^{\parallel}(x,t_{r})$ and
$\psi_{c}^{\parallel}(x,t_{r}+\frac{T}{2})$ from Eq.\ref{eq15} and
the symmetry relation between
$\tilde{\psi_{c}}^{\parallel}(k,t_{r})$ and
$\tilde{\psi_{c}}^{\parallel}(k,t_{r}+\frac{T}{2})$ from
Eq.\ref{eq16} still hold since the modification of the bound part of
the electronic wavefunction has no influence on the continuum part.

Using the facts that $\frac{\partial \phi_{1}(\mathbf{r})}{\partial
x}$ and $\nabla V_{0}(\mathbf{r})$ are antisymmetric with respect to
inversion of $x$, the following symmetry is obtained from
Eq.\ref{eq23}:

\begin{equation}
\tilde{\tilde{\psi}}^{(1)}(k,t_{r})=+\tilde{\tilde{\psi}}^{(1)}(-k,t_{r}+T/2)e^{-\frac{i}{\hbar}(I_{p}+E_{k})\frac{T}{2}}
\label{eq24}
\end{equation}

Using this symmetry, by calculating $\mathbf{a}^{(1)}(t_{r}+T/2)$
using the definition given in Eq.\ref{eq22}, together with the
relation given in Eq.\ref{eq24}, we get that:

\begin{equation}
\mathbf{a}^{(1)}(t_{r}+T/2)= +\mathbf{a}^{(1)}(t_{r}) \label{eq25}
\end{equation}

%\begin{eqnarray}
%\nonumber && \mathbf{a}_{IR}^{Stark}(t_{r}+T/2) \\
%\nonumber && = \Re \int_{-\infty}^{\infty}dk \tilde{\tilde{\psi}}_{IR}^{Stark}(k,t_{r}+T/2) \varepsilon_{1}^{out}
%\biggl\{e^{-\frac{i}{\hbar}[I_{p}+\frac{\hbar^{2}k^{2}}{2m}+\hbar\omega](t_{r}+T/2)}+e^{-\frac{i}{\hbar}[I_{p}+\frac{\hbar^{2}k^{2}}{2m}-\hbar\omega](t_{r}+T/2)}\biggr\} \\
%\nonumber && = \Re \int_{-\infty}^{\infty}dk \tilde{\tilde{\psi}}_{IR}^{Stark}(-k,t_{r})e^{+\frac{i}{\hbar}(I_{p}+E_{k})\frac{T}{2}} \varepsilon_{1}^{out}
%\biggl\{e^{-\frac{i}{\hbar}[I_{p}+\frac{\hbar^{2}k^{2}}{2m}+\hbar\omega](t_{r}+T/2)}+e^{-\frac{i}{\hbar}[I_{p}+\frac{\hbar^{2}k^{2}}{2m}-\hbar\omega](t_{r}+T/2)}\biggr\} \\
%\nonumber && = \Re \int_{-\infty}^{\infty}dk \tilde{\tilde{\psi}}_{IR}^{Stark}(-k,t_{r})e^{+\frac{i}{\hbar}(I_{p}+E_{k})\frac{T}{2}} \varepsilon_{1}^{out}
%\biggl\{e^{-\frac{i}{\hbar}[I_{p}+\frac{\hbar^{2}k^{2}}{2m}+\hbar\omega](t_{r}+T/2)}+e^{-\frac{i}{\hbar}[I_{p}+\frac{\hbar^{2}k^{2}}{2m}-\hbar\omega](t_{r}+T/2)}\biggr\} \\
% \label{eq33}
%\end{eqnarray}

The acceleration vector $\mathbf{a}^{(1)}(t_{r})$ is periodic in
$T/2$, i.e., its only nonzero components in its Fourier expansion
correspond to even integer harmonics of $\omega$. The term which is
responsible for the emission, $\varepsilon_{1}^{out}\cos(\omega
t_{r})\mathbf{a}^{(1)}(t_{r})$, however switches signs every $T/2$,
therefore giving rise to odd harmonics in the HGS ($\cos\alpha
\cos\beta =\frac{1}{2}[\cos(\alpha -\beta) + \cos(\alpha +\beta)]$).
That is, the radiation resulting from the ac-Stark oscillations of
the bound electron is composed of odd integer harmonics of $\omega$,
like the radiation which results from $\mathbf{a}^{(0)}$. The two
fields, emitted by $\mathbf{a}^{(0)}$ and
$\varepsilon_{1}^{out}\cos(\omega t_{r})\mathbf{a}^{(1)}$, interfere
with each other in general. However, since the field resulting from
$\mathbf{a}^{(1)}$ is much weaker, it is completely masked by the
field produced from $\mathbf{a}^{(0)}$.

The effect of the ac-Stark oscillations on the HGS could be
summarized as follows: In a single re-collision event each electron
trajectory (plane wave) with kinetic energy $E_{k}$, recombines with
the nucleus to emit radiation at energy $I_{p}+E_{k}$, and also, due
to the ac-Stark effect, two weaker "duplicate" electromagnetic waves
with energies $I_{p}+E_{k}-\hbar\omega$ and
$I_{p}+E_{k}+\hbar\omega$, \textit{at the same emission time}. In a
multi re-collision sequence, due to the symmetry properties
discussed above, only odd-harmonic photons will appear in the HGS.
The contribution to each plateau odd harmonic $\Omega$ in the HGS
comes, in principle, from six emission times: the first corresponds
to the recombination of the short trajectory with kinetic energy
$\hbar\Omega-I_{p}$, the second corresponds to the "duplicate"
recombination resulting from a different short trajectory, with
kinetic energy $\hbar\Omega-\hbar\omega-I_{p}$ (the ac-Stark
oscillations of the ground state at frequency $\omega$ will make the
final energy of the emitted photon
$\hbar\Omega-\hbar\omega-I_{p}+\hbar\omega=\hbar\Omega-I_{p}$) and
the third corresponds to the "duplicate" recombination resulting
from a different short trajectory, with kinetic energy
$\hbar\Omega+\hbar\omega-I_{p}$. Three additional emission times
result from three long trajectories in the same manner. Because of
the large differences in intensities, usually only the two "usual"
emission times attributed to the short and long trajectories at
kinetic energy $\hbar\Omega-I_{p}$ will contribute.

The effect of the ac-Stark oscillations on the HGS in this example
is not large, since the harmonics produced by this mechanism are
completely masked. However, high enough frequency of the ac-Stark
oscillations, well above the IR cut-off, will cause the appearance
of high energy photons which were not present at all in the IR HGS.
In this case, the lack of contribution of the ordinary re-collision
mechanism (neglecting the ac-Stark effect) to the appearance of
these new high-energy photons makes the ac-Stark effect the only
important one.

The way to induce ac-Stark oscillation of high frequency is by
applying a second high-frequency XUV field, in addition to the IR
one. The emphasis is that in order to see the new harmonics produces
by the "duplicate" trajectories due to the ac-Stark effect, the
second field should be close to the IR cut-off or above, otherwise
the new harmonics will be masked by the already existing IR ones.

Suppose then that we shine the atom with an IR field of frequency
$\omega$ and an XUV field of frequency $\tilde{q}\omega$. As
discussed before, the XUV field, provided that it has a large enough
frequency, doesn't affect the electron trajectories. Its only
influence is therefore on the recombination process. This field
induces ac-Stark oscillation in the exact same way as the IR field
did. We may assume the same approximations used before and
approximate the instantaneous ground state wavefunction as

\begin{equation}
\psi_{b}(\mathbf{r},t)\cong \phi_{1}(x+\varepsilon_{1}^{out}
\cos(\omega t)+\varepsilon_{\tilde{q}}^{out}
\cos(\tilde{q}\omega t),y,z)
e^{+\frac{i}{\hbar}I_{p}t} \label{eq26}
\end{equation}

The quiver amplitude $\varepsilon_{\tilde{q}}^{out}$ of the spatial
oscillations of the ground state is even smaller than
$\varepsilon_{1}^{out}$ because of the high frequency of the XUV
field (
$\varepsilon_{\tilde{q}}^{out}=\frac{\varepsilon_{\tilde{q}}^{in}}{(E_{2}+I_{p})^{2}-\tilde{q}^{2}\omega^{2}}\simeq
\frac{\varepsilon_{\tilde{q}}^{in}}{\tilde{q}^{2}\omega^{2}} $. The
bound part may be expanded in a Taylor serie as before

\begin{equation}
\psi_{b}(\mathbf{r},t)\cong e^{+\frac{i}{\hbar}I_{p}t}
\{\phi_{1}(\mathbf{r})+[\varepsilon_{1}^{out} \cos(\omega
t)+\varepsilon_{\tilde{q}}^{out} \cos(\tilde{q}\omega
t)]\frac{\partial}{\partial
x}\phi_{1}(\mathbf{r})\} \label{eq27}
\end{equation}
and the time-dependent acceleration expectation value is calculated
using the total wavefunction at the event of recombination
$\Psi(\mathbf{r},t\approx t_{r})$ with the modified bound part
$\psi_{b}(\mathbf{r},t_{r})$ given in Eq.\ref{eq27}. Keeping again
only the bound-continuum terms, we get:

\begin{equation}
\mathbf{a}(t_{r})= \mathbf{a}^{(0)}(t_{r})+[\varepsilon_{1}^{out}\cos(\omega t_{r})+\varepsilon_{\tilde{q}}^{out}\cos(\tilde{q}\omega t_{r})]\mathbf{a}^{(1)}(t_{r})
\label{eq28}
\end{equation}
where $\mathbf{a}^{(0)}(t_{r})$ is given in Eq.\ref{eq13} and
$\mathbf{a}^{(1)}(t_{r})$ in Eq.\ref{eq22}.

The ac-Stark effect at the frequency of the XUV field contributes a
new term $\varepsilon_{\tilde{q}}^{out}\cos(\tilde{q}\omega
t_{r})\mathbf{a}^{(1)}(t_{r})$ to the acceleration. In a single
re-collision event at $t\approx t_{r}$, this additional term
produces two bursts of light. One corresponds to the spectral
continuum
$I_{p}+\tilde{q}\hbar\omega<\hbar\Omega<I_{p}+3.17U_{p}+\tilde{q}\hbar\omega$
and the other to
$I_{p}-\tilde{q}\hbar\omega<\hbar\Omega<I_{p}+3.17U_{p}-\tilde{q}\hbar\omega$
which, in case that the XUV field is well above the IR cut-off,
could be written as
$\tilde{q}\hbar\omega-I_{p}<\hbar\Omega<\tilde{q}\hbar\omega-I_{p}-3.17U_{p}$.
These bursts of light are much weaker than the one which results
from $\mathbf{a}_{IR}(t_{r})$, nevertheless they have a significant
impact on the HGS since they are the only source of new XUV
harmonics which appear now in the HGS. As explained before, each
electron trajectory (plane wave) with kinetic energy $E_{k}$,
recombines with the nucleus to emit radiation of energy
$I_{p}+E_{k}$, and at the same time two "duplicate" photons with
energies
$I_{p}+E_{k}-\tilde{q}\hbar\omega=\tilde{q}\hbar\omega-I_{p}-E_{k}$
and $I_{p}+E_{k}+\tilde{q}\hbar\omega$.

When each single re-collision event is repeated every half cycle of
the IR field, the new term in the acceleration contributes to the
HGS integer even harmonics around $\tilde{q}$, i.e. $\tilde{q} \pm
2K$. To see this we compare two consecutive re-collision events at
times $t_{r}$ and $t_{r}+\frac{T}{2}$ and note that the result
obtained in Eq.\ref{eq25} still holds here:
$\mathbf{a}^{(1)}(t_{r}+T/2)= +\mathbf{a}^{(1)}(t_{r})$. The
acceleration vector $\mathbf{a}^{(1)}(t_{r})$ is periodic in $T/2$,
i.e., contributes even integer harmonics of $\omega$. The term which
is responsible for the emission,
$\varepsilon_{\tilde{q}}^{out}\cos(\tilde{q}\omega
t_{r})\mathbf{a}^{(1)}(t_{r})$ therefore gives rise to the
appearance of the harmonics $\Omega=(\tilde{q} \pm 2K)\omega$ in the
HGS ($\cos\alpha \cos\beta =\frac{1}{2}[\cos(\alpha -\beta) +
\cos(\alpha +\beta)]$). That is, the radiation resulting from the
ac-Stark oscillations of the bound electron is composed of even
integer harmonics of $\omega$ around the harmonic $\tilde{q}$. This
field is much weaker than the harmonics produced by
$\mathbf{a}^{(0)}$. In case that $\tilde{q}$ is an odd integer, the
term $\varepsilon_{\tilde{q}}^{out}\cos(\tilde{q}\omega
t_{r})\mathbf{a}^{(1)}(t_{r})$ will produce odd harmonics, which
will be masked by the IR harmonics produced from $\mathbf{a}^{(0)}$,
provided that $\tilde{q}$ is well below the IR cut-off harmonic.
However, in case that $\tilde{q}$ is close to or above the IR
cut-off harmonic or is not an odd integer, new harmonics (defined
before as "XUV harmonics"), which were not present in the HGS in the
presence of the IR field alone, will appear. These harmonics will be
$|\varepsilon_{\tilde{q}}^{out}|^{2} \simeq
(\frac{\varepsilon_{\tilde{q}}^{in}}{\tilde{q}^{2}\omega^{2}})^{2}
$-times weaker than the IR harmonics, and could therefore be
distinguished from the IR harmonics by their intensity. Since these
harmonics are originated by the same electronic trajectories which
produce the IR harmonics, their emission times are correlated with
the ones of the IR harmonics. In a single re-collision event each
electron trajectory (plane wave) with kinetic energy $E_{k}$,
recombines with the nucleus to emit radiation at energy
$I_{p}+E_{k}$, and also, due to the ac-Stark effect, two weaker
"duplicate" electromagnetic waves with energies
$I_{p}+E_{k}-\tilde{q}\hbar\omega$ and
$I_{p}+E_{k}+\tilde{q}\hbar\omega$, \textit{at the same emission
time}. This correlation is kept also in the multi re-collision
process and is manifested in the HGS: the structure (amplitude and
phase) of the XUV harmonics (XUV-HGS) $\Omega=(\tilde{q} \pm
2K)\omega$ is derived from the structure of the IR-HGS. This is true
for every value of $\tilde{q}$ but could be most easily seen if
$\tilde{q}$ is well above the IR cut-off harmonic, since in this
case the XUV harmonics are separated and are not nested in the
IR-HGS. If we look at the case $\tilde{q}=52$ in Fig.\ref{fig1}, we
see that the structure of the XUV-HGS between the orders $54$ and
$74$ resembles the structure of the IR-HGS between the orders 1 and
21 ($54-52-1=1$ , $74-52-1=21$). In addition, within the XUV-HGS,
the structure of harmonics between the orders $32$ and $50$ is a
mirror-image (with respect to the $52$-nd harmonics) of the
structure of harmonics between the orders $54$ and $72$. That is,
the XUV-HGS consists of two new plateau-like regions (harmonics of
order 38-50 and 54-66), derived from the same electronic
trajectories which form the IR-HGS plateau (harmonics of order
1-15), and two new cut-off-like regions (harmonics of order 32-36
and 68-74), derived from the same electronic trajectories which form
the IR-HGS cut-off harmonics (harmonics of order 17-23). This
structure of the XUV-HGS is invariant to the value of $\tilde{q}$.

What happens if we now take higher-order terms in the Taylor serie
expansion of the ground state wavefunction in Eq.\ref{eq26}? Suppose
we take also the second-order term in the Taylor serie and expand
the bound state according to

\begin{equation}
\psi_{b}(\mathbf{r},t)\cong e^{+\frac{i}{\hbar}I_{p}t}
\{\phi_{1}(\mathbf{r})+[\varepsilon_{1}^{out} \cos(\omega
t)+\varepsilon_{\tilde{q}}^{out} \cos(\tilde{q}\omega
t)]\frac{\partial}{\partial
x}\phi_{1}(\mathbf{r})+[\varepsilon_{1}^{out} \cos(\omega
t)+\varepsilon_{\tilde{q}}^{out} \cos(\tilde{q}\omega
t)]^{2}\frac{1}{2}\frac{\partial^{2}}{\partial
x^{2}}\phi_{1}(\mathbf{r})\} \label{eq29}
\end{equation}

The time-dependent acceleration expectation value, keeping again
only the bound-continuum terms, would read:

\begin{eqnarray}
\nonumber && \mathbf{a}(t_{r})= \mathbf{a}^{(0)}(t_{r})+[\varepsilon_{1}^{out}\cos(\omega t_{r})+\varepsilon_{\tilde{q}}^{out}\cos(\tilde{q}\omega t_{r})]\mathbf{a}^{(1)}(t_{r})+[\varepsilon_{1}^{out}\cos(\omega t_{r})+\varepsilon_{\tilde{q}}^{out}\cos(\tilde{q}\omega t_{r})]^{2}\mathbf{a}^{(2)}(t_{r}) \\
\nonumber && =\mathbf{a}^{(0)}(t_{r})+[\varepsilon_{1}^{out}\cos(\omega t_{r})+\varepsilon_{\tilde{q}}^{out}\cos(\tilde{q}\omega t_{r})]\mathbf{a}^{(1)}(t_{r}) \\
\nonumber && + \biggl\{ (\varepsilon_{1}^{out})^{2}\biggl[\frac{1+\cos(2\omega t_{r})}{2} \biggr] +
\frac{1}{2}\varepsilon_{1}^{out}\varepsilon_{\tilde{q}}^{out}\biggl[
\cos[(\tilde{q}+1)\omega t_{r}]+\cos[(\tilde{q}-1)\omega t_{r}]
\biggr] + (\varepsilon_{\tilde{q}}^{out})^{2} \biggl[
\frac{1+\cos(2\tilde{q}\omega t_{r})}{2} \biggr] \biggr\}
\mathbf{a}^{(2)}(t_{r}) \\
\nonumber && =\mathbf{a}^{(0)}(t_{r})+[\varepsilon_{1}^{out}\cos(\omega t_{r})+\varepsilon_{\tilde{q}}^{out}\cos(\tilde{q}\omega t_{r})]\mathbf{a}^{(1)}(t_{r}) \\
&& + \biggl\{ \frac{(\varepsilon_{1}^{out})^{2}+(\varepsilon_{\tilde{q}}^{out})^{2}}{2}+\frac{(\varepsilon_{1}^{out})^{2}}{2}\cos(2\omega t_{r})
+\frac{\varepsilon_{1}^{out}\varepsilon_{\tilde{q}}^{out}}{2}\cos[(\tilde{q}+1)\omega t_{r}]+\frac{\varepsilon_{1}^{out}\varepsilon_{\tilde{q}}^{out}}{2}\cos[(\tilde{q}-1)\omega t_{r}]
+ \frac{(\varepsilon_{\tilde{q}}^{out})^{2}}{2} \cos(2\tilde{q}\omega t_{r})\biggr\}
\mathbf{a}^{(2)}(t_{r}) \label{eq30}
\end{eqnarray}
where $\mathbf{a}^{(0)}(t_{r})$ is given in Eq.\ref{eq13},
$\mathbf{a}^{(1)}(t_{r})$ in Eq.\ref{eq22}, and
$\mathbf{a}^{(2)}(t_{r})$ is

\begin{equation}
\mathbf{a}^{(2)}(t_{r}) = \Re \int_{-\infty}^{\infty} dk \tilde{\tilde{\psi}}^{(2)}(k,t_{r}) e^{-\frac{i}{\hbar}[I_{p}+\frac{\hbar^{2}k^{2}}{2m}]t_{r}}
\label{eq31}
\end{equation}
where

\begin{equation}
\tilde{\tilde{\psi}}^{(2)}(k,t_{r})\equiv -\frac{1}{m}\frac{1}{\sqrt{2\pi}}
\tilde{\psi_{c}}^{\parallel}(k,t_{r})\int_{-\infty}^{\infty} d^{3}r
\frac{\partial^{2} \phi_{1}(\mathbf{r})}{\partial x^{2}}\nabla V_{0}(\mathbf{r})
\psi_{c}^{\perp}(y,z,t_{r})e^{ikx} \label{eq32}
\end{equation}

The new term $+[\varepsilon_{1}^{out}\cos(\omega
t_{r})+\varepsilon_{\tilde{q}}^{out}\cos(\tilde{q}\omega
t_{r})]^{2}\mathbf{a}^{(2)}(t_{r})$, which results from the
inclusion of the second-order term in the Taylor serie expansion of
the bound wavefunction, produces 10 weaker "duplicate" bursts of
light in a single recollision event at $t\approx t_{r}$. This is due
to the fact that the term multiplying $\mathbf{a}^{(2)}(t_{r})$ has
5 different frequency components. These bursts of light are much
weaker than the one which results from $\mathbf{a}^{(0)}(t_{r})$ or
$+[\varepsilon_{1}^{out}\cos(\omega
t_{r})+\varepsilon_{\tilde{q}}^{out}\cos(\tilde{q}\omega
t_{r})]\mathbf{a}^{(1)}(t_{r})$. Nevertheless, it will be shown
immediately that the last two bursts (resulting from
$\frac{(\varepsilon_{\tilde{q}}^{out})^{2}}{2} \cos(2\tilde{q}\omega
t_{r}) \mathbf{a}^{(2)}(t_{r})$) have a significant impact on the
HGS since they are the only source of new XUV harmonics which appear
around the harmonic $2\tilde{q}$.

Let us analyze now what happens when each single re-collision event
is repeated every half cycle of the IR field. It was shown in
Eq.\ref{eq18} that the term $\mathbf{a}^{(0)}(t_{r})$ contributes
odd harmonics $\Omega=(2K-1)\omega$ to the HGS. From Eq.\ref{eq25}
the term $\varepsilon_{1}^{out}\cos(\omega
t_{r})\mathbf{a}^{(1)}(t_{r})$ also contributes odd harmonics
$\Omega=(2K-1)\omega$ (with relative amplitude of the electric field
of $\varepsilon_{1}^{out}$) but the term
$\varepsilon_{\tilde{q}}^{out}\cos(\tilde{q}\omega
t_{r})\mathbf{a}^{(1)}(t_{r})$ contributes the XUV harmonics
$\Omega=(\tilde{q} \pm 2K)\omega$. Under the assumption that
$\phi_{1}(\mathbf{r})$ is symmetric, $\frac{\partial^{2}
\phi_{1}(\mathbf{r})}{\partial x^{2}}$ is also symmetric. Therefore,
in complete analogy to what was shown in Eq.\ref{eq18}, it can be
shown that the term $\mathbf{a}^{(2)}(t_{r})$ contributes odd
harmonics of $\omega$ since in two consecutive re-collision events
at times $t_{r}$ and $t_{r}+\frac{T}{2}$ the following symmetry
holds:

\begin{equation}
\mathbf{a}^{(2)}(t_{r}+T/2)= -\mathbf{a}^{(2)}(t_{r}) \label{eq33}
\end{equation}

The term which is responsible for the emission,
$[\varepsilon_{1}^{out}\cos(\omega
t_{r})+\varepsilon_{\tilde{q}}^{out}\cos(\tilde{q}\omega
t_{r})]^{2}\mathbf{a}^{(2)}(t_{r})$ therefore gives rise to the
appearance of the following five sets of harmonics (with the
amplitudes given in parentheses): $\Omega=[0 \pm
(2K-1)]\omega=(2K-1)\omega$
[$(\varepsilon_{1}^{out})^{2}+(\varepsilon_{\tilde{q}}^{out})^{2}$],
$\Omega=[2 \pm (2K-1)]\omega=(2K-1)\omega$
[$(\varepsilon_{1}^{out})^{2}$], $\Omega=[\tilde{q}+1 \pm
(2K-1)]\omega=(\tilde{q} \pm 2K)\omega$
[$\varepsilon_{1}^{out}\varepsilon_{\tilde{q}}^{out}$],
$\Omega=[\tilde{q}-1 \pm (2K-1)]\omega=(\tilde{q} \pm 2K)\omega$
[$\varepsilon_{1}^{out}\varepsilon_{\tilde{q}}^{out}$],
$\Omega=[2\tilde{q} \pm (2K-1)]\omega$
[$(\varepsilon_{\tilde{q}}^{out})^{2}$]. Out of these five sets, the
first four contribute harmonics which are masked, due to their low
intensity, by IR harmonics or by the XUV harmonics resulting from
the lower-order term
$\varepsilon_{\tilde{q}}^{out}\cos(\tilde{q}\omega
t_{r})\mathbf{a}^{(1)}(t_{r})$. The fifth set, however, contribute a
new set of harmonics: $\Omega=[2\tilde{q} \pm (2K-1)]\omega$. This
set is $|(\varepsilon_{\tilde{q}}^{out})^{2}|^{2} \simeq
(\frac{\varepsilon_{\tilde{q}}^{in}}{\tilde{q}^{2}\omega^{2}})^{4}
$-times weaker than the IR harmonics, or
$|\varepsilon_{\tilde{q}}^{out}|^{2}$-times weaker than the set
$\Omega=(\tilde{q} \pm 2K)\omega$ of XUV harmonics and could
therefore be distinguished from these set of harmonics by their
intensity. As before, these harmonics are originated by the same
electronic trajectories which produce the IR harmonics, and their
emission times are correlated with the ones of the IR harmonics.
This is manifested in the HGS: the structure (amplitude and phase)
of the new set of XUV harmonics $\Omega=[2\tilde{q} \pm
(2K-1)]\omega$ in the XUV-HGS is derived from the structure of the
IR-HGS. It consists, like the set $\Omega=(\tilde{q} \pm 2K)\omega$,
of two new plateau-like regions, derived from the same electronic
trajectories which form the IR-HGS plateau, and two new cut-off-like
regions, derived from the same electronic trajectories which form
the IR-HGS cut-off harmonics. This structure of the XUV-HGS is
almost invariant to the value of $\tilde{q}$, as can be easily seen
from Fig.\ref{fig2}.

\begin{figure}[ht]
\hbox{\includegraphics[width=3.8in,height=2.8in]{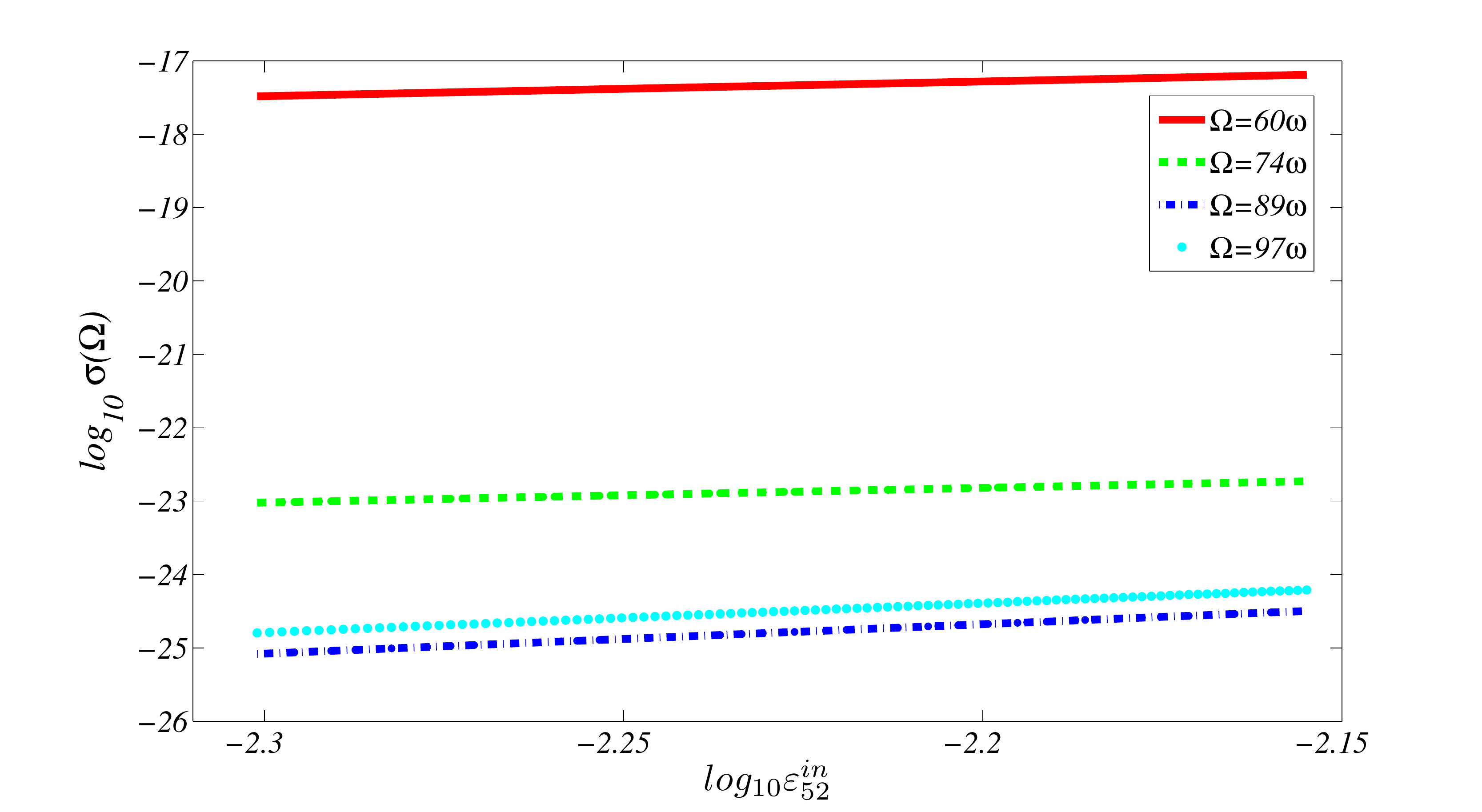}}%[width=3.8in,height=2.8in]
\caption{\label{fig5}\small (color online) log-log plot of the intensity of different harmonics $\sigma(\Omega)$ [$\Omega=60$ (solid red line), $\Omega=74$ (dashed green line),
$\Omega=89$ (dash-dotted blue line), $\Omega=97$ (dotted cyan line)] obtained from the 1D
model Hamiltonian of Xe atom (Eq.\ref{eq3}) irradiated by a 50-oscillation
sine-square pulse of bichromatic laser field composed of a 800nm
IR laser field of intensity $I^{in}_{1}\simeq 4.299\cdot 10^{13}W/cm^{2}$ and a $800/52-$nm XUV
field, as function of the XUV field's amplitude $\varepsilon^{in}_{52}$. This intensity varies between $0.005a.u.$
($I^{in}_{52}\simeq 8.773\cdot 10^{11}W/cm^{2}$) and $0.007a.u.$ ($I^{in}_{52}\simeq 1.719\cdot 10^{12}W/cm^{2}$).
All 4 graphs are linear. According to the lower frame of Fig.\ref{fig2} ($\tilde{q}=52$), the harmonics $\Omega=60$ and $\Omega=74$ belong to the set $\tilde{q} \pm 2K$ of XUV harmonics,
and as such should depend on $\varepsilon^{in}_{52}$ as $\sigma(\Omega=\tilde{q} \pm 2K)\propto (\varepsilon^{in}_{52})^{2}$ (see Eq.\ref{eq28}).
Indeed the slopes of the graphs corresponding to $\Omega=60$ and $\Omega=74$ are $2.000$ and $2.0001$, respectively. The harmonics $\Omega=89$ and $\Omega=97$ belong to the set $2\tilde{q} \pm (2K-1)$ of XUV harmonics,
and as such should depend on $\varepsilon^{in}_{52}$ as $\sigma[\Omega=2\tilde{q} \pm (2K-1)]\propto (\varepsilon^{in}_{52})^{4}$ (see Eq.\ref{eq30}).
Indeed the slopes of the graphs corresponding to $\Omega=89$ and $\Omega=97$ are $4.0045$ and $4.0015$, respectively.
This check confirms the correctness of our theory in this range of parameters.}
\end{figure}

By generalizing the procedure described before and taking
higher-order terms in the Taylor serie expansion of the bound
wavefunction, it is apparent that the n-th order term contributes a
new set of XUV harmonics $\Omega=[n\tilde{q} \pm
(2K-1+mod(n,2))]\omega$ which is
$|(\varepsilon_{\tilde{q}}^{out})^{n}|^{2} \simeq
(\frac{\varepsilon_{\tilde{q}}^{in}}{\tilde{q}^{2}\omega^{2}})^{2n}
$-times weaker than the IR harmonics. Note that indeed in
Fig.\ref{fig2} for $\tilde{q}=52$, the intensity of the plateau
harmonics of the set $\Omega=(\tilde{q} \pm 2K)\omega$ is indeed
$|\varepsilon_{\tilde{q}}^{out})|^{2} \simeq
(\frac{\varepsilon_{\tilde{q}}^{in}}{\tilde{q}^{2}\omega^{2}})^{2}=(\frac{0.0035}{52^{2}0.05695^{2}})^{2}\simeq
1.6\cdot 10^{-7} $-times weaker than the IR plateau harmonics, and
the intensity of the plateau harmonics of the set
$\Omega=[2\tilde{q} \pm (2K-1)]\omega$ is indeed
$|\varepsilon_{\tilde{q}}^{out})|^{4} \simeq 2.5\cdot 10^{-14}
$-times weaker than the IR plateau harmonics, in agreement with our
theory. Fig.\ref{fig5}, which plots the intensity of different
harmonics as function of the amplitude
$\varepsilon_{\tilde{q}}^{in}$ of the XUV driver field, shows indeed
that the intensity of harmonics from the set $\Omega=(\tilde{q} \pm
2K)\omega$ scale quadratically with $\varepsilon_{\tilde{q}}^{in}$,
while harmonics from the set $\Omega=[2\tilde{q} \pm (2K-1)]\omega$
scale as $(\varepsilon_{\tilde{q}}^{in})^{4}$. The obtained sets of
XUV harmonics $\Omega=[n\tilde{q} \pm (2K-1+mod(n,2))]\omega$ are
exactly as predicted by the selection-rules given in Eq.\ref{eq7}.
Hence, the inclusion of higher-order terms in the Taylor serie
expansion of the bound wavefunction, leads to the generalization of
the semiclassical three-step model. This allows us to obtain the
selection rules for the high harmonics which are obtained upon the
addition of an XUV field to an IR one, which are in complete
agreement with the ones obtained using Floquet theory \cite{Avner
Atto}. Moreover, the intensities of the XUV harmonics can also be
quantified.

%60: y=2x-12.882   74:y=2.0063x-18.404  89: y=4.1428x-15.548
%97:y=3.9151x-15.784

%176-276:   60: y=2x-12.882   74:y=2.0001x-18.419  89: y=4.0045x-15.866  97:y=4.0015x-15.585

\begin{figure}[ht]
\hbox{\includegraphics[width=3.8in,height=2.8in]{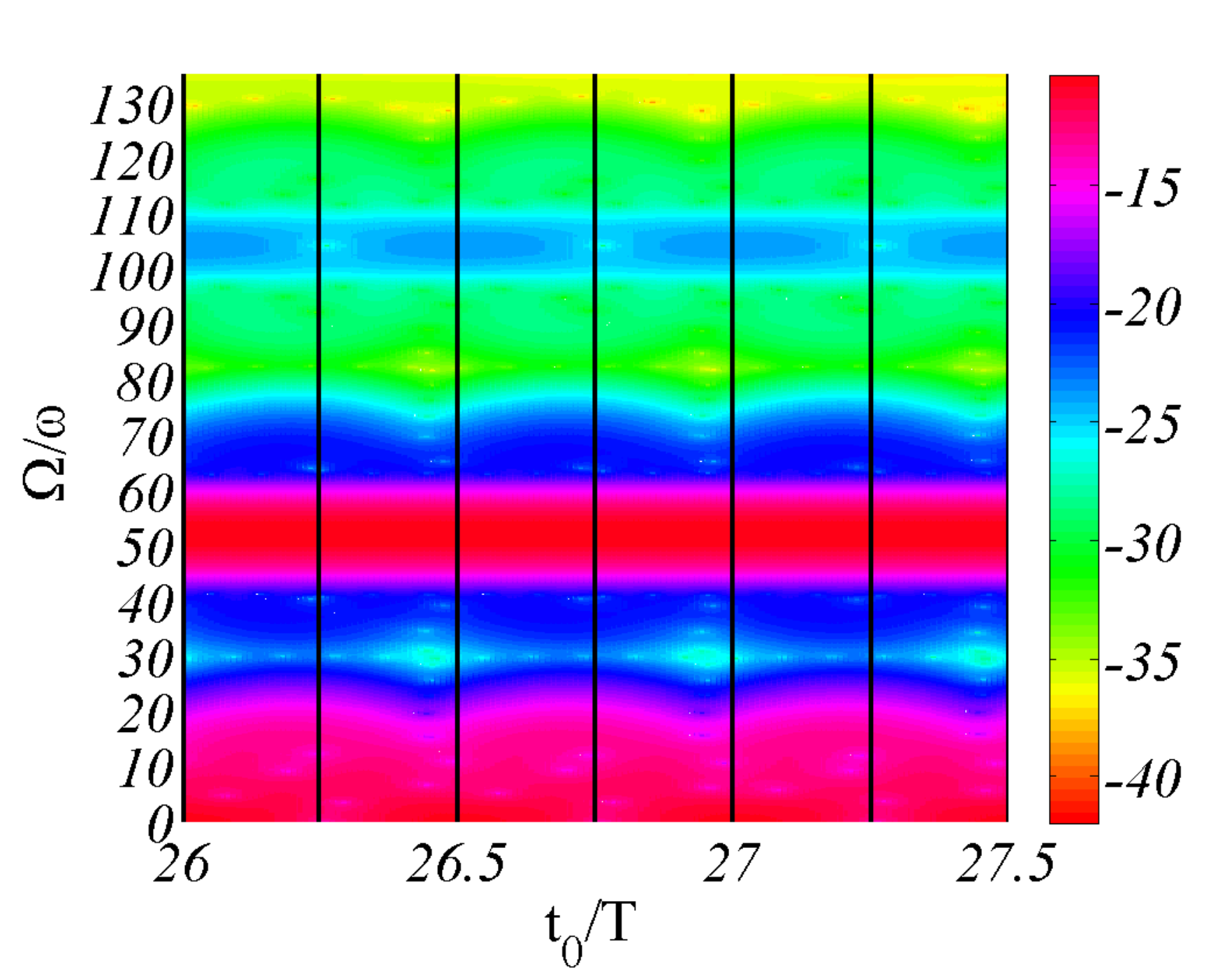}}
\caption{\label{fig6}\small (Color online) Top view [pink (dark gray) color—high
intensity, yellow (bright gray) color—low intensity] of the absolute square
of the Gabor-tansformed acceleration expectation value (
Eq.\ref{eq34}, $\tau=0.1T$) of the quantum mechanical simulation described
in Fig.\ref{fig2} for $\tilde{q}=52$, as function of $t_{0}$ and $\Omega$.}
\end{figure}

The above theory predicts that both the IR and the XUV harmonics are
emitted by the same IR trajectories and are therefore correlated in
their emission times. This could be easily verified by plotting the
time-frequency distribution of high harmonics, i.e., by analyzing
the Gabor-transform (windowed Fourier transform) of the acceleration
instead of the Fourier transform:

\begin{equation}
G_{a}(\Omega,t_{0})=\frac{1}{NT}\int_{0}^{NT}
a(t)e^{-\frac{(t-t_{0})^{2}}{\tau^{2}}} e^{-i\Omega t } dt \label{eq34}
\end{equation}
where $\tau$ is the window's width. This analysis yields a "mixed"
time-frequency signal, i.e., not only the frequency components
appearing in the acceleration but also their time of appearance
$t_{0}$. Fig.\ref{fig6} shows the time-frequency distribution of
high harmonics $(\sigma(\Omega,t_{0})\equiv \frac{2e^{2}}{3c^{3}}
|G_{a}(\Omega,t_{0})|^{2})$ obtained from the time-dependent
acceleration expectation value whose spectra is given in
Fig.\ref{fig2} for $\tilde{q}=52$, for the times $26T < t < 27.5T$.
In accordance with the semiclassical re-collision model, different
harmonics are emitted repeatedly every half cycle. In accordance
with the theory developed here, the time-frequency distribution of
the new sets of XUV harmonics matches the one of the IR harmonics,
including the reflection symmetry of these sets (around $\tilde{q},
2\tilde{q}...$). The most visible demonstration of this property is
shown for the IR cut-off harmonics (the 15th-29th harmonic), who are
emitted at times $t\simeq 0.710T +0.5nT$, in accordance with the
semiclassical re-collision model. At those instants, also the
32nd-36th and the 68th-74th harmonics, which are the XUV cut-off
harmonics of the set $\tilde{q} \pm 2K$, and also the 85th-87th and
the 121st-125th harmonics, which are the XUV cut-off harmonics of
the set $2\tilde{q} \pm (2K-1)$, are emitted. They are hence
produced by the IR cut-off trajectories.

\section{Suggested Experiment and Conclusions}

The mechanism described here raises the question whether XUV
harmonics should be self-produced in any monochromatic HHG
experiment, as high-harmonic radiation generated by the leading edge
of the IR pulse, co-propagates with the IR field to form a
bichromatic driver field in the last part of the medium. The answer
is that the high-harmonic radiation generated in the medium is too
weak to considerably modify the IR-HGS. If, on the other hand, a
stronger high-frequency source is used, and in particular with
frequency higher than the IR cut-off frequency, new XUV harmonics
should appear. This source could be a free-electron laser, but could
also be a HHG-based source. Generation of HHG pulses of the 27th
harmonic of Ti:sapphire laser, with width of 30fs and output energy
of $0.33 \mu J$ per pulse, has been shown to be feasible in Ar
\cite{Takahashi}. When such an harmonic pulse is focused to an area
of $(10 \mu m)^{2}$, intensities of $I^{in}_{27}\simeq
10^{13}W/cm^{2}$ may be reached. It isn't unreasonable to assume
that higher-harmonics couldn't be generated with similar output
energies. Even a reduction of 5 orders of magnitude in the intensity
of the high-harmonic field will make the effect still visible. For
example, by generating the 45th harmonic in Xe or He, filtering it
out of the HGS and focusing it into a jet of Kr together with an IR
field that is sufficient to generate IR cut-off at the 19th harmonic
or so, new high harmonics should appear. Kr has a spectral window
between photon energies of $50eV$ (the 31st harmonic) and $90eV$
(the 57th harmonic). Therefore, shining a 45th-harmonic field on it
will not cause single-XUV photon ionization and will cause the
appearance of new XUV harmonics in this spectral window, without the
necessity to increase the intensity of the IR field.

Alternatively, in case that the frequency of the high-harmonic field
isn't large enough, one can reduce the intensity of the IR field in
order to decrease the IR cut-off. The key point here is that in
order to see the effect of appearance of new XUV harmonics the XUV
field should be strong enough and of frequency higher than the IR
cut-off frequency. Above all, the gas in which the bichromatic HHG
is generated, should be transparent in some spectral band around the
frequency of the XUV field, otherwise the generated XUV harmonics
and/or the seed XUV field will be absorbed.

In conclusion, we have shown that the addition of an XUV field to a
strong IR field leads to the appearance of new harmonics in the HGS.
The results of the semiclassical analysis and the quantum numerical
simulations suggest that this is a single-atom phenomena. The XUV
field induces ac-Stark modulations on the ground state and affects
the recombination process of all returning trajectories, and leads
to the generation of higher harmonics whose emission times and
intensities are well related to the ones of the harmonics in the
presence of the IR field alone. Using this mechanism, harmonics with
unprecedented high frequencies could be obtained in HHG experiments.
According to our mechanism, the emitted HHG radiation field could be
written as a serie of terms, with the zeroth-order term equal the
HHG field which is obtained from the three-step model in its most
familiar context \cite{P. B. Corkum}. The higher-order terms become
important when, in addition to the IR field, an additional XUV field
is shined on the atom since they solely are responsible for the
appearance of new harmonics in the HGS.

This work was supported in part by the by the Israel Science
Foundation and by the Fund of promotion of research at the Technion.
We thank the second referee of \cite{Avner Rapid} whose comments
triggered us to develop the mechanism described in this paper.

\bibliographystyle{plain}

\end{document}